\documentclass[prb,twocolumn,showpacs,preprintnumbers,amsmath,amssymb]{revtex4}
\usepackage{graphicx}
\usepackage{dcolumn}
\usepackage{bm}
\begin{document}
\title {Study of electron spin dynamics in grain aligned LaCoPO: an itinerant ferromagnet}
\author{M. Majumder$^1$}
\author{K. Ghoshray$^1$}
\email{kajal.ghoshray@saha.ac.in}
\author{A. Ghoshray$^1$}
\author{B. Bandyopadhyay$^1$}
\author{M. Ghosh$^2$}

\affiliation{$^1$ECMP Division, Saha Institute of Nuclear Physics, 1/AF Bidhannagar,
Kolkata-700064, India}
\affiliation{$^2$Hooghly Women's College, Pipulpati, Hooghly}
\date{\today}
\begin{abstract}

$^{139}$La NMR study was performed in grain aligned ($\overrightarrow{c}$$\parallel$$\overrightarrow{H_0}$) sample of LaCoPO and polycrystalline LaFePO. Knight shift is isotropic and temperature independent in LaFePO. It is strongly temperature dependent and anisotropic in LaCoPO. The spin-lattice relaxation rate in LaCoPO clearly reveals the existence of 3D spin fluctuations both in the paramagnetic and ferromagnetic state over and above the dominant 2D spin fluctuations in the paramagnetic state, observed earlier from $^{31}$P NMR measurements in the same oriented sample\cite{Majumder09}. The spin fluctuation parameters in LaCoPO determined from $^{139}$La NMR relaxation and magnetization data, using the self consistent renormalization (SCR) theory, are in close agreement and follow the universal Rhodes-Wohlfarth curve.
\end{abstract}
\pacs{74.70.-b, 76.60.-k}
\maketitle

\section{INTRODUCTION}

Materials having strongly correlated electrons exhibit exotic magnetic and electronic properties including superconductivity with high transition temperature,\cite{Ishida09} itinerant ferromagnetism, giant magnetoresistance etc. due to the existence of several competing states. Recently discovered layered compounds $LnT_{\mathrm{M}}PnO$ [$Ln \equiv 4f$ rare earth element, $T_\mathrm{M}$=transition metal element, $Pn$=pnictogen element] have attracted attention of the condensed matter physicists for their various interesting properties when doped\cite{Kamihara08} with electron or hole or being pressurized.\cite{Takahashi08} Like cuprates, pnictides have also two dimensional electronic structure where the  crystal structure is composed of alternate stacks of $Ln$O and $T_{\mathrm{M}}Pn$ layers. The magnetic properties of La$T_{\mathrm{M}}$PO are of special interest with respect to the variation of $T_{\mathrm{M}}$ i.e. changing the number of 3$d$ electrons. In case of Fe (3$d^6$) and Ni (3$d^8$) as $T_{\mathrm{M}}$, the oxypnictides undergo superconducting transition (with $T_\mathrm{c}$ lying in the range of 3-6 K), where the magnetic ordering is suppressed due to reduction of magnetic moments.\cite{Hirano08,Tegel07} For Zn (3$d^{10}$) and Mn (3$d^5$) as $T_{\mathrm{M}}$, the compounds show semiconducting and antiferromagnetic semiconducting properties respectively.\cite{Kayanuma07,Yanagi09} In LaCoPO, the magnetic moment does not vanish completely due to the odd number of electrons in the 3$d$ orbitals (3$d^7$), thereby  showing ferromagnetic transition at 43 K, when the magnetization is measured in an external magnetic field of 0.1 T, with no superconducting transition down to 2 K.\cite{Yanagi08} Thus it is emerged that the number of 3$d$ electrons have a definite influence on the macroscopic properties of these type of compounds.

In the present paper, we report the results of $^{139}$La NMR studies in a grain aligned sample of LaCoPO in which $^{31}$P NMR \cite{Majumder09,Sugawara09} showed the presence of dominant ferromagnetic (FM) 2D spin fluctuations in the paramagnetic
state. $^{31}$P being a spin 1/2 nucleus, possesses no quadrupole moment, therefore it provides information only about the dominant magnetic interactions within the Co-P plane. The non magnetic La-O plane is situated in between the two magnetic Co-P planes (Fig. 1\label{structure}), with Co and O atoms sitting at the center of the P and La tetrahedra respectively. Therefore, the $^{139}$La NMR Knight shift and the spin-lattice relaxation time
measurements are expected to provide additional information about the microscopic nature of the magnetism (compared to that obtained from
$^{31}$P NMR) by probing different portion of the electronic structure. Furthermore, La being a quadrupolar nucleus, the nature of the local electronic charge distribution through the electrostatic hyperfine interaction parameters would be revealed. We have analyzed the magnetization data in LaCoPO using the self consistent renormalization (SCR) theory of
Takahashi\cite{Takahashi86} and determined the spin fluctuation parameters which are then compared with that determined from $^{139}$La NMR
spin-lattice relaxation data again using SCR theory.\cite{Moriya85,Ishigaki96} We also present the results of $^{139}$La NMR studies in the
polycrystalline sample of LaFePO (which does not order magnetically and exhibit superconducting transition $\simeq$ 4 K,\cite{Kamihara08}) in order to compare the effect of the substitution of Fe in place of Co, on the electronic properties in such
compounds.
\begin{figure}[h]
{\centering {\includegraphics{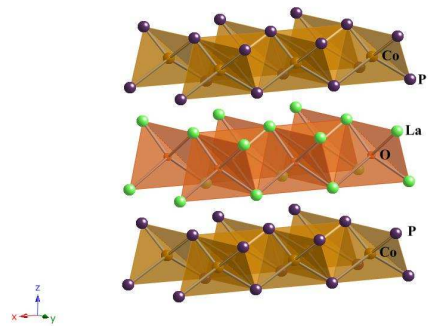}}\par}
\caption{(Color online) Crystal structure of tetragonal (space group P4/nmm) LaCoPO consists of alternating La-O and Co-P layers arranged along the $z$-axis.}
\label{structure}
\end{figure}

\section{EXPERIMENTAL}

Polycrystalline samples of LaFePO and LaCoPO were synthesized by the method of solid state reaction~\cite{Yanagi08,Hirano08} and were characterized using powder x-ray diffraction technique with CuK$\alpha$ radiation at room temperature. Almost all the diffraction peaks were assigned within the space group P4/nmm. Weak peaks with intensity $\sim$ 4\% with respect to the strongest diffraction peak of LaCoPO were assigned as due to La$_2$O$_3$, present as an impurity phase. The magnetizations ($M$) were measured as functions of temperature ($T$) and the magnetic fields ($H$)  using a SQUID magnetometer (MPMSXL 7 T, Quantum Design). The NMR measurements were carried out in a conventional phase-coherent spectrometer (Thamway PROT 4103MR) with a 7.04 T($H_0$) superconducting magnet from Bruker. The spectrum was recorded by changing the frequency step by step and recording the spin echo intensity by applying a $\pi/2-\tau-\pi/2$ solid echo sequence. The temperature variation in the range 4 - 300 K was performed in an Oxford continuous flow cryostat with a ITC503 controller. The powder sample of LaCoPO was aligned in the NMR coil, by mixing the fine powders with an epoxy (Epotek-301) and keeping overnight in the magnetic field of 7.04 T.

\section{RESULTS AND DISCUSSIONS}

\subsection{Measurement of magnetization}

The variation of magnetization ($M$) with temperature in a magnetic field of $H$=0.1 T shows a sharp enhancement below 50 K due to ferromagnetic ordering, as reported earlier.\cite{Yanagi08,Majumder09} Derivative of the $M$ vs $T$ curve provides an approximate value of $T_\mathrm{C}$ to be 35 K (inset of Fig. 2). Fig. 2 shows the isothermal $M$ versus $H$ curves in the range 2-60 K. Above $H$=2 T, the curves show a saturation in the low temperature region, and around 60 K the $M$ versus $H$ curve becomes linear, indicating that the system reached the  paramagnetic state.
For homogeneous itinerant ferromagnet
\begin {equation}
M(T,H)^2 = M_s(T)^2 + \zeta.H/M(T,H),
\end {equation}
where the coefficient $\zeta$ is independent of temperature around $T_\mathrm{c}$. The $T_\mathrm{c}$ is defined as the temperature at which the spontaneous magnetization ($M_{\mathrm{s}}$) is zero. From the Arrott plot ($M(T,H)^2$ vs $H/M(T,H)$ plots) (Fig. 3) one can estimate $T_\mathrm{c}$. It is seen that the curves are convex in nature similar to that of itinerant ferromagnet LaCoAsO\cite{Ohta09}. This nature of the curves was predicted by Takahashi's SCR theory of spin fluctuations,\cite{Takahashi86}
where it is assumed that the sum of the zero point and the thermal spin fluctuations is conserved against $T$.
\begin{figure}[h]
{\centering {\includegraphics{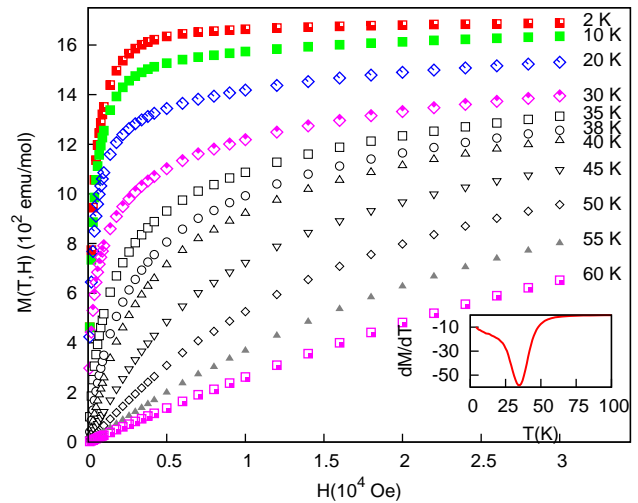}}\par}
\caption{(Color online) $M$ vs $H$ curve for polycrystalline LaCoPO at different temperatures. Inset shows the $dM/dT$ vs $T$ curve for $H$ = 0.1 $T$.}
\label{susceptibility}
\end{figure}

\begin{figure}[h]
{\centering {\includegraphics{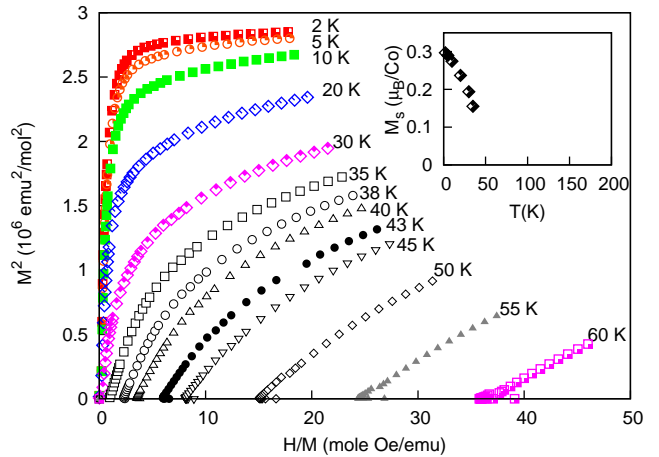}}\par}
\caption{(Color online) Arrott plot for LaCoPO. In the inset $M_s$ vs $T$ curve is shown.}
\label{susceptibility}
\end{figure}
\begin{figure}[h]
{\centering {\includegraphics{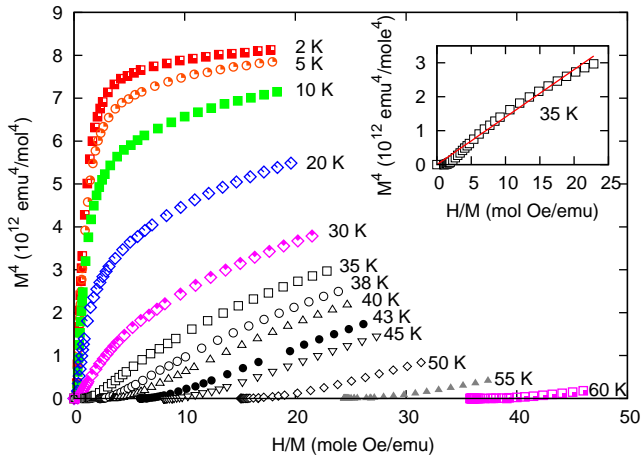}}\par}
\caption{(Color online) $M^4$ vs $H/M$ is plotted and in the inset $M^4$ vs $H/M$ curve at $T_{\mathrm{C}}$ (35 K) is plotted and the solid line is the linear fit.}
\label{susceptibility}
\end{figure}
\begin{figure}[h]
{\centering {\includegraphics{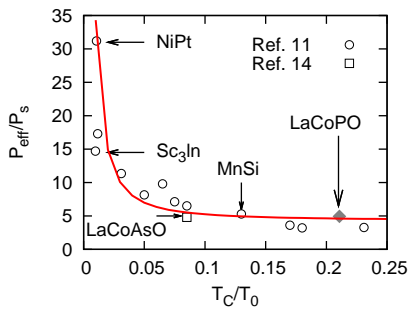}}\par}
\caption{(Color online) $P_{\mathrm{eff}}/P_{\mathrm{s}}$ vs $T_\mathrm{C}$/$T_0$ plot (Rhodes-Wohlfarth plot). Solid line corresponds to Rhodes-Wohlfarth equation ie. $P_{\mathrm{eff}}/P_{\mathrm{s}}$ $\propto$ ($T_\mathrm{C}$/$T_0$)$^{-3/2}$.}
\label{susceptibility}
\end{figure}

In the localized limit, zero-point spin fluctuation is negligible and the thermal spin fluctuation is constant above $T_\mathrm{C}$. However, in case of weakly itinerant ferromagnetic metals, the zero-point spin fluctuation mainly forms the total spin fluctuation and the thermal spin fluctuation increases with temperature above $T_\mathrm{C}$, which gives the value of the ratio of the effective paramagnetic moment to the saturation moment, $M_{\mathrm{eff}}/M_{\mathrm{s0}}$ greater than 1, which is $\sim$ 1, in case of localized ferromagnet. From the $M_{\mathrm{s}}$ vs $T$ curve (inset of Fig. 3) the approximate value of $M_{\mathrm{s0}}$, the value of $M_{\mathrm{s}}$ at $T$=0 is obtained as 0.3 $\mu_B$ per Co ion, which is close to the reported value.\cite{Yanagi08} $P_{\mathrm{eff}}$ ($M_{\mathrm{eff}}$ in $\mu_B$ unit) estimated from the slope of Curie-Weiss plot in the range 300 -70 K is 1.49 $\mu_B$. This is greater than $M_{\mathrm{s0}}$ signifying the itinerant character of the local spin moments.

Fig. 4 shows isothermal magnetization curves in the form of $M^4$ vs $H/M$ for LaCoPO. In itinerant ferromagnet, $M^4$ follows
the relation\cite{Takahashi86,Ohta09}
\begin {equation}
M^4 = 1.17\times 10^{19} ({T_C}^2/{T_A}^3)(H/M)
\end {equation}
at $T_\mathrm{C}$, where $T_\mathrm{A}$ characterizes the width of the distribution of the static susceptibility in the wave vector ($q$) space.
$M^4$ vs $H/M$ curve  at $T_\mathrm{C}$ (inset of Fig. 4) shows not much departure from a linear fitting (indicated by the continuous line), and therefore closely supports the Takahashi's theory suggesting a linear behavior represented by Eq. (2), when the coefficient of $M^4$ term in the Landau expansion of free energy becomes zero. The value of $T_A$ (shown in Table I) is estimated from the slope of $M^4$ vs $H/M$ curve at $T_\mathrm{C}$. According to the SCR thoery of weak itinerant ferromagnet (WIF)\cite{Ohta09,Moriya73,Takahashi85,Moriya85}
\begin {equation}
T_C = (60c)^{-3/4}P_s^{3/2}T_A^{3/4}T_0^{1/4}
\end {equation}
where $c$=0.3353$\cdot\cdot\cdot$ and $P_{\mathrm{s}}$ is $M_{\mathrm{s0}}$ in $\mu_B$ unit. $T_0$ which characterizes the energy width of the dynamical spin-fluctuation spectrum is obtained (Table I) from Eq. 3. This  theory is also used to evaluate the coefficient $\bar{F}_1$ (an important spin fluctuation parameter) of $M^4$ term, in the Landau expansion of free energy, which can be written as\cite{Takahashi86}
\begin {equation}
\bar{F}_1 = 4T_A^2/15T_0.
\end {equation}
$\bar{F}_1$ as estimated (Table I) from Eq. 4 is
found to be larger than that reported for LaCoAsO.\cite{Ohta09} This suggests a more localized character of the $d$-electrons in LaCoPO compared to that in LaCoAsO.

We also verified that LaCoPO follows universal Rhodes-Wohlfarth relation\cite{Rhodes63} between $P_{\mathrm{eff}}/P_{\mathrm{s}}$ and $T_\mathrm{C}$/$T_0$ viz., $P_{\mathrm{eff}}/P_{\mathrm{s}}\sim(T_\mathrm{C}/T_0)^{-3/2}$ as shown in Fig. 5.
According to the SCR theory, $T_\mathrm{C}$/$T_0$ = 1 in case of localized ferromagnet, but as the itinerant character sets in, the value becomes less than one. In case of LaCoAsO, the value of $T_\mathrm{C}$/$T_0$ $<$ 0.1.\cite{Ohta09} Whereas, it is $>$ 0.2 for LaCoPO as obtained from the present results. This also suggests a more localized character of the d-electrons in LaCoPO than in LaCoAsO. Table I shows the obtained values of the parameters from the present magnetization study in LaCoPO and those reported for LaCoAsO\cite{Ohta09} for comparison.

\begin{table}[h]
\caption{Parameters deduced from the present magnetization data in LaCoPO and those reported for LaCoAsO\cite{Ohta09}.}
\begin{tabular}{l l l}
\hline \hline
$\textrm{Parameters}$ & LaCoPO  &  LaCoAsO\\
\hline
$T_\mathrm{C}$ (K) & $35$ & $55$\\
$T_\mathrm{A}$ (K) & $4606$ & $6210$\\
$T_0$ (K) & $166$ & $644$\\
$\bar{F}_1$ (K) & $34080$ & $16000$\\
$P_{\mathrm{eff}}$ ($\mu_B$) & $1.49$ & $1.34$\\
$P_{\mathrm{s}}$ ($\mu_B$) & $0.3$ & $0.28$\\
$P_{\mathrm{eff}}$/P$_{\mathrm{s}}$ & $4.9$ & $4.8$\\
$T_\mathrm{C}$/$T_0$ & $0.211$ & $0.085$\\
\hline\hline
\end{tabular}
\end{table}

\subsection{$^{139}$La NMR Knight shift and hyperfine coupling constant in LaCoPO }

\begin{figure}[h]
{\centering {\includegraphics{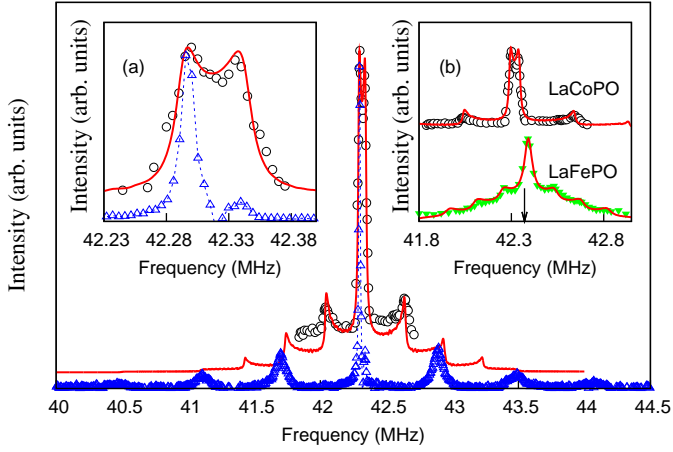}}\par}
\caption{(Color online) $^{139}$La NMR spectra in LaCoPO at 7.04 T: powder pattern($\bigcirc$) with central and first pair of satellite lines along with the theoretical fit (solid line), central line with all satellites in aligned sample ($\bigtriangleup$). Inset (a): zoom part of central transition for powder($\bigcirc$) and aligned sample($\bigtriangleup$) in LaCoPO. Inset (b): powder pattern of $^{139}$La NMR spectra in LaFePO (central and all satellites) and LaCoPO (central and first pair of satellites) along with the solid line representing the theoretical spectrum, $\downarrow$ indicates reference frequency for $^{139}$La.}
\label{susceptibility}
\end{figure}

$^{139}$La ($I$=7/2) NMR study was performed in the same grain aligned sample of LaCoPO, in which the $^{31}$P NMR results were reported
previously.\cite{Majumder09} Fig. 6 shows the NMR spectrum corresponding to the central and the six satellite transitions, appearing due to the existence of non-zero electric field gradient (EFG) at the La nuclear site, in grain aligned sample of LaCoPO at 300 K, along with the spectrum recorded in for the polycrystalline sample showing only the central and the pair of first satellite lines. It is seen that the separation between the first satellites in grain aligned sample is twice than that of the same in random powder. This shows that the grains in the aligned sample are oriented in a direction of $\overrightarrow{c}$$\parallel$$\overrightarrow{H_0}$ i.e., $\theta$=0, where $\theta$ represents the angle between the direction of the external magnetic field ($H_0$) and the z-principal axis of the electric field gradient (EFG) tensor,\cite{Cohen57} which in this case lies along the crystallographic c-axis. The continuous superimposed line represents the theoretical spectrum and is generated by considering the quadrupolar and magnetic hyperfine interactions as perturbations over the Zeeman term. Moreover it was assumed that the principal axes of the EFG and the magnetic shift tensors are parallel to each other.\cite{Carter77}

The low frequency peak of the central transition in the polycrystalline sample (shown more clearly in the inset (a) of Fig. 6) aries due to the anisotropy of the magnetic hyperfine/dipolar interaction, with the z-principal axis of the magnetic hyperfine interaction tensor parallel to the external magnetic field ($\theta$=0). Therefore, the matching of the position of the central transition of the aligned sample with this peak corresponding to the polycrystalline sample, further confirms that the grains in the aligned sample are oriented along the $\theta$=0 direction, in agreement with that observed from the satellite separation. Therefore, it justifies the assumption that the principal axes of the EFG and the magnetic hyperfine interaction tensors are coincident. For aligned sample, the small peak [inset (a)] at the high frequency side corresponds to the signal $\theta$=$\pi$/2, indicating the presence of a small portion of unaligned part as was also seen in $^{31}$P NMR of the same sample.\cite{Majumder09}

Inset (b) of Fig. 6 shows the $^{139}$La NMR in polycrystalline LaFePO along with that of LaCoPO at 300 K. In both the cases, the theoretical line shape (represented by the solid line) is superimposed on the experimental line. Clearly, the internal magnetic field at the La nuclear site in LaFePO is isotropic, whereas it is axially symmetric in LaCoPO. Moreover, the electric field gradient (EFG) at the La nuclear site is smaller in LaFePO than in LaCoPO, as also revealed from the satellite pair separation. This can happen if the contribution of the non $s$ electronic orbitals to the conduction band is enhanced because of Co ion, resulting in the increment in EFG as well as lowering the symmetry of the local magnetic field at the La nuclear site. On the other hand, replacement of P by As increases the EFG at the La site, but enhances the symmetry of the local magnetic field\cite{Ohta10} (Table II). Thus it is emerged that the detailed knowledge of the band structure showing the involvement of the La electronic orbitals in the conduction band is necessary to understand these findings.

\begin{figure}
{\centering {\includegraphics{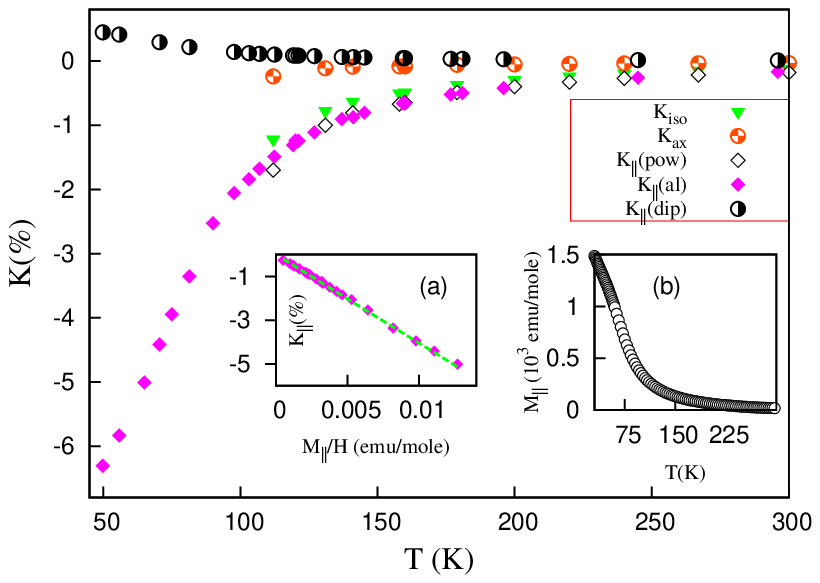}}\par}[h]
\caption{(Color online) $K_\|$(\%) vs $T$(K) for $^{139}$La NMR in aligned LaCoPO($\blacklozenge$). $T$ dependence of $K_{\mathrm{iso}}$(\%), $K_{\mathrm{ax}}$(\%), $K_\|$(\%) determined from powder pattern and calculated $K_\|(\mathrm{dip})$(\%) vs $T$ curve. Inset (a): $K_\|$(\%) vs $M_\|/H$ curve for $^{139}$La NMR in aligned LaCoPO. The solid line is the linear fit. Inset (b): $M_\|$ vs $T$(K) curve at 7 T.}
\label{susceptibility}
\end{figure}
Figure 7 shows the temperature dependence of the $^{139}$La NMR Knight shift $K_\|$(\%) measured in the aligned sample in the temperature range 50-300 K. $K_\|$(\%) corresponds to the Knight shift, measured with the z-principal axis of the hyperfine coupling tensor (in the present case it is the crystallographic $c$-axis), parallel to the direction of the external magnetic field, $H_0$. $K_\perp$ is the Knight shift measured along the principal axis which is perpendicular to the direction of $H$ (in the present case it is either $a$ or $b$ axis). The variation of $K_{\textrm{iso}}$$(\%)$, $K_{\textrm{ax}}$$(\%)$ and $K_\|$($\%$), in the temperature range of 110-300 K, obtained by theoretically fitting the NMR spectra of the random powder sample, are also shown in the same figure. The isotropic part of the Knight shift $K_{\textrm{iso}}$ = 2$K_\perp$/3 + $K_\|$/3 and the axial part $K_{\textrm{ax}}$ = 1/3($K_\|$ - $K_\perp$).  The inset (b) of Fig. 7 shows the magnetization $M_\|$ versus $T$ curve in the aligned sample in the range 4-300 K for $H$=7 Tesla.

In case of $d$-band metals, the bulk magnetic susceptibility $\chi$ can be written as
\begin {equation}
\chi(T) = \chi_{\textrm{dia}} + \chi_{\textrm{orb}} + \chi_\textrm{s} + \chi_\textrm{d}(T)
\end {equation}
where $\chi_d(T)$ and $\chi_{orb}$ are the spin and the orbital susceptibilities of $d$ electrons per Co atom, $\chi_s$ is the spin susceptibilities
of $s$ and $p$ conduction electrons, and $\chi_{dia}$ is the diamagnetic susceptibilities of core electrons. Only $\chi_d(T)$ gives a temperature
dependent contribution to the measured susceptibility. Similarly, the measured shift can be written as
\begin {equation}
K = K_0 + K_\textrm{d}(T)
\end {equation}
where $K_0$ contains the sum of the contributions due to $\chi_{orb}$, $\chi_s$ and $\chi_{dia}$. $K_d(T)$ contains the contribution due to the $d$-electron spin and can be written as,
\begin {equation}
K_\textrm{d}(T) = (H_{\textrm{hf}}/N\mu_B)\chi_d(T)
\end {equation}
where $H_{\textrm{hf}}$ is the sum of the hyperfine and the dipolar field, $N$ is the Avogadro number and $\mu_B$ is the Bohr magneton.
The inset (a) of Fig. 7 shows the variation of $K_\|$($\%$) with $M_\|/H$ in the temperature range 65-300 K. The linear variation of the intrinsic spin susceptibility, probed by the NMR shift with the bulk susceptibility, indicates that the macroscopic magnetization contains negligible contribution from the magnetic impurity phase, and arises from the electronic moments homogeneously distributed within the sample. Moreover, the straight line passes almost through the origin, indicating a negligible contribution of $K_0$ to the total Knight shift of $^{139}$La NMR in LaCoPO similar to that reported in LaCoAsO.\cite{Ohta10} From Fig. 1 it seems there is negligible overlap between the orbitals of La and Co. Therefore, the temperature dependent $^{139}$La NMR Knight shift could arise from the hyperfine field transferred from the Co-3$d$ spins, through the conduction electron bands by Rudderman-Kittel-Kasuya (RKKY) type exchange interaction to the La nuclear site, together with the magnetic dipolar field produced by the magnetic moment of the Co atoms situated at the different lattice points. In Fig. 7 we have also shown the $T$ dependence of $K_\|(\mathrm{dip})$, calculated using the method of lattice sum. It is seen that the dipolar contribution to the shift for $\overrightarrow{c}$$\parallel$$\overrightarrow{H_0}$ is positive and its magnitude increases slowly with the lowering of $T$. This shows that the negative sign of the experimental shift with sharp enhancement in magnitude below 150 K, originates from purely hyperfine contribution produced by a negative exchange interaction between the La-6s conduction electrons and the Co-3d electrons.

Table II shows the values of $H_{\mathrm{hf}}^{\|}$, $H_{\mathrm{hf}}^{\mathrm{iso}}$ and $H_{\mathrm{hf}}^{\mathrm{ax}}$ for LaCoPO obtained from appropriate form of Eq. 7. The magnitude of all the parameters in LaCoPO are found to be almost twice of those reported in LaCoAsO by Ohta et al.\cite{Ohta10} The $^{139}$La NMR Knight shift (0.04\%) in LaFePO is almost temperature independent in the range 4-300 K, whereas $^{31}$P Knight shift in LaFePO\cite{Majumder09} and in Ca-doped LaFePO\cite{Nakai08} showed a weak T dependence similar to that of $\chi($T$)$. Therefore, it is emerged that in the $^{139}$La NMR Knight shift of LaFePO, $K_0$ is the dominant contribution and $K_d$($T$) arising from Fe-3$d$ electrons is negligible. Because of the layered crystal structure, the small magnetic moment in the Fe-3$d$ orbital, could not polarize the La-$s$ electrons to produce a significant magnitude of $K_d$($T$) in the La NMR Knight shift. Whereas, P being directly bonded with the Fe atom, it could even sense the weak T dependence of $\chi_d$($T$)\cite{Nakai08} through the magnetic hyperfine interaction.

\begin{table}[h]
\caption{Values of magnetic hyperfine coupling constants ($H_\mathrm{hf}$) and the quadrupolar splitting frequency ($\nu_Q$) obtained from $^{139}$La NMR in LaCoPO and those reported in LaCoAsO.\cite{Ohta10}}
\begin{tabular}{l l l}
\hline \hline
$\textrm{Parameters}$ &  LaCoPO &  LaCoAsO\cite{Ohta10}\\
\hline
$H_{\mathrm{hf}}^{\|}$ (kOe/$\mu_B$) &  $-22.6$ &  $-11.4$\\
$H_{\mathrm{hf}}^{\mathrm{iso}}$ (kOe/$\mu_B$) &  $-17.58$ &  $-8.64$\\
$H_{\mathrm{hf}}^{\mathrm{ax}}$ (kOe/$\mu_B$) &  $-2.23$ &  $-1.41$\\
$\nu_Q$ (MHz) &   $0.57$ &   $1.48$\\
\hline\hline
\end{tabular}
\end{table}

\subsection{ Nuclear spin-lattice relaxation rate 1/$T_1$ }

Fig. 8 shows (1/$T_1T$)$_{H\|c}$ versus temperature curve represented by the open triangles, for $^{139}$La nucleus in LaCoPO in the temperature range 30 - 300 K. Appearance of a peak at around 106 K, indicates that the La nucleus which is situated in the La-O layer, separated from the magnetic Co-P layer, could also sense the effect of the development of the ferromagnetic correlation among the Co-3d spins in LaCoPO.
In general, (1/$T_1T$)$_{\mathrm{SF}}$ is given by
\begin {equation}
(1/T_1T)_{\mathrm{SF}} = 2\gamma_n^2k_B\sum_\textbf{q}H_{\mathrm{hf}}(\textbf{q})^2\chi\prime\prime(\textbf{q},\omega_n)/\omega_n,
\end {equation}
where $\chi\prime\prime(q,\omega_n$) is the imaginary part of the transverse dynamical electron-spin susceptibility, $\gamma_n$ and $\omega_n$ are the nuclear gyromagnetic ratio and Larmor frequency, respectively. $H_{\mathrm{hf}}(\textbf{q})$ = $\Sigma_j H_{\mathrm{hf}}^j \exp(-i\textbf{q}.\textbf{r}^j)$ at the La site and $H_{\mathrm{hf}}^j$ is the hyperfine coupling constant between the La site and the nearest neighbor Co electron spins and $\textbf{r}^j$ is the position vectors of Co sites from the La site. Since for weakly ferromagnetic metals, only the small $\textbf{q}$ components are important in the $\textbf{q}$-summation, one can replace $H_{\mathrm{hf}}(\textbf{q})$ by $H_{\mathrm{hf}}$(0) and it can be taken out of the summation. Therefore, Eq. (8) can be rewritten as
\begin {equation}
(1/T_1T)_{\mathrm{SF}} = 2\gamma_n^2k_BH_{\mathrm{hf}}(0)^2\sum_\textbf{q}\chi\prime\prime(\textbf{q},\omega_n)/\omega_n,
\end {equation}

\begin{figure}[h]
{\centering {\includegraphics{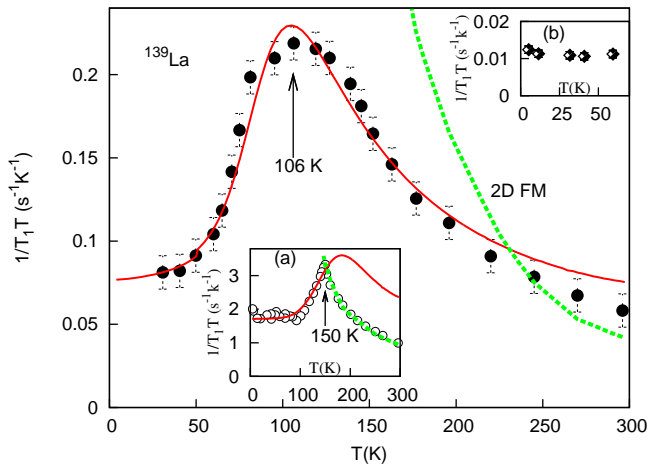}}\par}
\caption{(Color online) (1/$T_1T)_{H\|c}$ vs $T$(K) curve for $^{139}$La NMR in LaCoPO. The continuous line corresponds to theoretical curve according to Eq. 12. Inset (a): variation of (1/$T_1T)_{H\|c}$ with respect to $T$ for $^{31}$P NMR in LaCoPO, continuous line corresponds to Eq. 12, broken line corresponds to Eq. 13. Inset (b): variation of 1/$T_1T$ with respect to temperature for $^{139}$La NMR in LaFePO.}
\label{susceptibility}
\end{figure}
In general, 1/$T_1T$ is decomposed as
\begin {equation}
1/T_1T = (1/T_1T)_\textrm{d} + (1/T_1T)_{\textrm{orb}} + (1/T_1T)_\textrm{s} + (1/T_1T)_{\textrm{dip}}
\end {equation}
where $(1/T_1T)_{\textrm{orb}}$ and $(1/T_1T)_{\textrm{dip}}$ represent orbital moment contributions from $p$ and $d$ electrons and spin-dipolar interaction with $p$ and $d$ electrons respectively, $(1/T_1T)_\textrm{s}$ represents Fermi-contact contribution of $s$ conduction electrons and $(1/T_1T)_\textrm{d}$ represents the hyperfine contributions of $d$ electrons. Assuming $(1/T_1T)_s$ and $(1/T_1T)_{orb}$ are temperature independent and neglecting $(1/T_1T)_{\textrm{dip}}$ with respect to $(1/T_1T)_\textrm{d}$, as the dipolar coupling constant($H_{\textrm{dip}}^\|$=1.47 kOe/$\mu_B$ as evaluated from lattice sum) is negligibly small with respect to hyperfine coupling constant ($H_{\textrm{hf}}^\|$ in Table II), $(1/T_1T)_d$ will be the only temperature dependent term.

According to the SCR theory for weak itinerant ferromagnetic (WIF) materials, where $(1/T_1T)_\textrm{d}$ is governed by 3D spin fluctuations of $d$ electrons far above $T_\mathrm{C}$,\cite{Ishigaki98,Yoshimura88,Yoshimura87,Nowak09}
\begin {equation}
(1/T_1T) = k\chi_d + \beta,
\end {equation}
where $\beta$ is the temperature independent contributions in Eq. 10 and k$\chi_d$ is $(1/T_1T)_d$. In Fig. 9 we have plotted $(1/T_1T)_{H\|c}$ versus $M_\|/H$, in the temperature range 120-300K, which clearly shows the validity of Eq. 11 in the paramagnetic state, with the continuous line representing the equation (1/$T_1T)_{H\|c}$ = 16$\times$10$^3$ $\chi_\|$ + 0.045.
It is therefore clear that in the paramagnetic region 3D spin-fluctuations govern the $^{139}$La nuclear relaxation process.
In the ordered state, where the magnetization is not linear in magnetic field, 1/$T_1T$ should have a field dependence. According to SCR theory of spin fluctuations, for WIF, the ferromagnetic ($\textbf{q}$=0) 3D spin-fluctuation contribution to 1/$T_1T$ in presence of magnetic field, both in the paramagnetic and ferromagnetic region is given by\cite{Moriya85,Rabis05,Yoshimura87}
\begin {equation}
 (1/T_1T)_{\mathrm{SF}} = k\chi_d/(1+\chi_d^3H^2P) + \beta,
\end {equation}
where $P$ is a constant related to the area of the fermi surface of the magnetic electrons and $k$ and $\beta$ are same as in Eq. 11. In Fig. 8
we have also plotted the theoretical curve (represented by the continuous line) for (1/$T_1T)_{H\|c}$ at versus temperatures using eq. 12 and use $\chi_\|$ in place of $\chi_d$. It is seen that Eq. 12 fits quite
satisfactorily with almost same $k$ and $\beta$ values as used in Eq. 11 (Fig. 9). In this figure, we have also shown the theoretical curve (represented by the dashed line) for 1/$T_1T$ by using the equation\cite{Nakai08,Ishigaki98,Hatatani95}
\begin {equation}
1/T_1T \propto \chi(\textbf{q}=0)^{3/2}
\end {equation}
as predicted by the SCR theory for weak itinerant ferromagnet, when 2D spin-fluctuations govern the $T_1$ process. The calculated curve is found to deviate significantly from the  experimental data. Therefore, we conclude that the fluctuating magnetic field
experienced by the La nucleus, is predominantly 3D in nature, both in the paramagnetic as well as in the ferromagnetic region. The inset (a) of Fig. 8 shows the
$^{31}$P (1/$T_1T)_{H\|c}$ versus temperature curve both in the paramagnetic and the ordered state.\cite{Majumder09}
\begin{figure}[h]
{\centering {\includegraphics{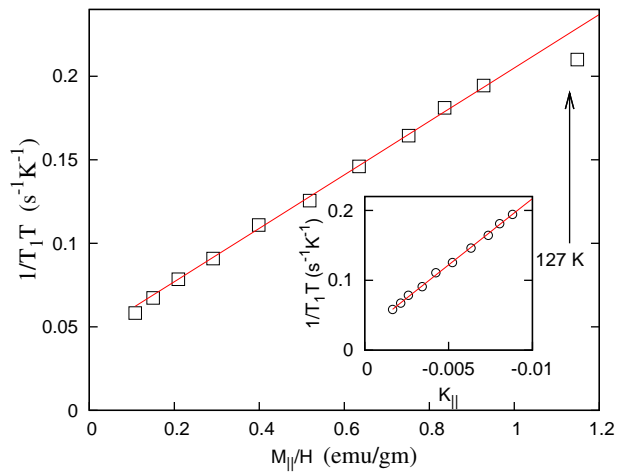}}\par}
\caption{(Color online) (1/$T_1T)_{H\|c}$ vs $M_\|/H$ for $^{139}$La NMR in LaCoPO and the solid line corresponds to Eq. 11. In the inset (1/$T_1T)_{H\|c}$ vs $K_\|$ for $^{139}$La NMR in LaCoPO and solid line corresponds to Eq. 17.}
\label{susceptibility}
\end{figure}
In this case, Eq. 12 fits well (using $\chi_\|$ of a sample not containing nonmagnetic impurity) only in the magnetically ordered region (represented by the continuous line) and deviates significantly in the paramagnetic region. The experimental data in the paramagnetic region is well fitted (dotted line) with equation eq.(13), suggesting also the presence of 2D spin fluctuations in the paramagnetic phase of LaCoPO. This observation clearly indicates that in the paramagnetic state, over and above the presence of dominant 2D-ferromagnetic spin fluctuations, there is also a non negligible contribution of the 3D spin fluctuations of Co$^{2+}$ 3$d$-electrons, through which the $^{139}$La nuclear relaxation rate, sense the effect of the development of the ferromagnetic ordering.

It is to be noted from Fig. 8 that for La NMR, the peak appears at a temperature which is about 40 K lower than that observed in $^{31}$P NMR relaxation. Also the value of $^{139}$(1/$T_1T$)/$\gamma_n^2$ is an order of magnitude smaller than that of $^{31}$(1/$T_1T$)/$\gamma_n^2$ in LaCoPO, similar to that reported in Ca doped LaFePO (which does not order magnetically) by Nakai et al\cite{Nakai08} from $^{31}$P and $^{139}$La NMR studies. These observations indicate that the magnitude of the 3D component in the ferromagnetic spin fluctuation is much weaker compared that of the 2D component in LaCoPO. One possible reason for this could be the layered structure, with the La atom being situated in a separate plane from the magnetic Co-P plane. Therefore, it could be expected that the La nuclear spin-lattice relaxation process could sense the existence of any 3D component in the over all spin fluctuation spectrum, arising due to the presence of the inter layer exchange interaction, in such layered compound. Therefore, a comparison of the $^{139}$La 1/$T_1$ data with those of $^{31}$P in the same grain aligned sample reveals the coexistence of stronger 2D ferromagnetic (FM) spin fluctuations along with comparatively weaker 3D FM
spin fluctuations in the paramagnetic state of itinerant ferromagnetic system LaCoPO.

$^{139}$La NMR $1/T_1T$ in LaFePO is almost temperature independent (Inset (b) of Fig. 8) indicating the dominating Korringa process in the relaxation mechanism and the contribution of Fe-3d spin fluctuation is negligible. Whereas, $^{31}$P relaxation in the same compound showed the presence of weak 2D ferromagnetic correlation of Fe-3d spins\cite{Majumder09} similar to that reported in Ca-doped LaFePO from NMR.\cite{Nakai08} This finding along with a similar behavior of Knight shift as discussed in Sec.IIIB, supports a quasi 2D electronic structure in such compounds.

\subsection{Spin fluctuation parameters from $^{139}$La nuclear relaxation data in LaCoPO}

Following Ishikagi and Moriya,\cite{Ishigaki98} it is possible to write $\chi\prime\prime(\textbf{q},\omega_n)/\omega_n$ in Eq. 8 in terms of two spin fluctuation parameters $T_0$ and $T_\mathrm{A}$ (defined in section III.A) which characterize the width of the spin excitations spectrum in frequency and wave vector ($\textbf{q}$) respectively.
For ferromagnetic correlations one can write\cite{Ishigaki98,Ishigaki96}
\begin {equation}
\chi(\textbf{q},\omega) = \frac{\pi T_0}{\alpha_QT_A}(\frac{x}{k_B2\pi T_0x(y+x^2)-i\omega\hbar}),
\end {equation}
where x=$\textbf{q}/q_B$ with $q_B$ being the effective zone boundary vector, $\alpha_Q$ a dimensionless interaction constant close to unity for a strongly correlated system, y = 1/2$\alpha_Qk_BT_A\chi(0,0)$. Here the susceptibility is per spin and in units of 4$\mu_\mathrm{B}^2$ and has the dimension of inverse of energy, $T_0$ and $T_\mathrm{A}$ are in Kelvin. From Eq. 9 one can derive $\chi\prime\prime(\textbf{q},\omega_n)$ in the limit $\omega_n$ $\rightarrow$ 0, since $\hbar\omega_n$ $\ll$ $k_BT$. For 3D spin fluctuations governing the relaxation process, one has to integrate $\chi\prime\prime(\textbf{q},\omega_n)/\omega_n$, over a sphere of radius $\textbf{q}_B$($(\frac{6\pi^2}{v_0})^{1/3}$),\cite{Corti07} whereas, in case of 2D spin fluctuations, the integration has to be done over a disc of radius $\textbf{q}_B$(($\frac{4\pi}{v_0})^{1/2}$). Therefore, in the former case we have,
\begin {equation}
\frac{1}{T_1} = \frac{\gamma^2H_{hf}^2}{2}T(\frac{3\hbar}{4\pi k_BT_AT_0\alpha_Q})(\frac{1}{2y(1+y)}).
\end {equation}
Here the reduced inverse susceptibility y = $\frac{1}{2T_\mathrm{A}\chi(0,0)}$. The measured susceptibility $\chi$ is related to $\chi$(0,0) as
$\chi$ = g$^2\mu_\mathrm{B}^2$$\chi(0,0)N_\mathrm{A}$. Since $T_\mathrm{A}$ $\gg$ $T$ ( as estimated from the magnetic susceptibility data shown in Table I) in the temperature range of the experiment, $y\ll$1. In this case the above equation can be simplified as
\begin {equation}
\frac{1}{T_1} \simeq \gamma^2H_{hf}^2\frac{3\hbar}{8\pi}(\frac{T}{T_0})\chi(0,0).
\end {equation}
Replacing $\chi$(0,0) by the Knight shift ($K$) = g$\mu_\mathrm{B}H_{\mathrm{hf}}$$\chi$(0,0) one can write,\cite{Corti07}
\begin {equation}
(1/T_1T) \simeq 3\hbar\gamma_n^2H_{hf}K/16\pi\mu_BT_0.
\end {equation}

\begin{table}[h]
\caption{Values of spin-fluctuation parameters deduced from magnetization data and $^{139}$La NMR data for LaCoPO.}
\begin{tabular}{l l l}
\hline \hline
$\textrm{Parameters}$ & Magnetization & NMR\\
\hline
$T_0$ (K) & $166$ & $115$\\
$T_\mathrm{A}$ (K) & $4606$ & $5207$\\
$\Gamma_0 (meV{\AA}^3)$ & $29.6$ & $20.5$\\
\hline\hline
\end{tabular}
\end{table}

Hence by plotting (1/$T_1T)_{H\|c}$ versus $K_\|$ (Inset of Fig. 9), we have calculated $T_0$ from its slope. $T_0$ is related to $\Gamma_0$, corresponding to the width in energy of the spin excitations\cite{Hatatani95} by,
\begin {equation}
T_0 = \Gamma_0q_B^3/2\pi
\end {equation}
where $q_B^3$ = $\frac{6\pi^2}{v_0}$  with $v_0$ = 19.49 ${\AA}^3$, corresponding to the atomic volume of Co. The parameter $T_\mathrm{A}$ is then calculated using Eq. 3. Spin fluctuation parameters evaluated from the $^{139}$La NMR results are given in Table III along with the same determined from the magnetization data. It is to be noted that the values agree in order of magnitude. In case of MnSi, another weakly itinerant magnet, the estimated value of $T_0$ from inelastic neutron scattering\cite{Ishikawa85} is 230 K and the same estimated from magnetic contribution to the thermal expansion coefficient\cite{Matsunaga82} is 231 K, showing a close agreement. However, the value of $T_0$ obtained from $^{29}$Si NMR 1/$T_1$ data\cite{Corti07} is 71 K which is one order of magnitude smaller. Even in the present case also, the value of $T_0$ obtained from $^{139}$La NMR in LaCoPO, is smaller than that obtained from the magnetization data, though they are of same order of magnitude. This finding possibly supports the argument of Corti et al\cite{Corti07} regarding such disagreement that in the expression for $q_B^3$, $v_0$ should be the primitive cell volume and not Co atomic volume, with $q_B$ close to the zone boundary.
SCR theory for weak itinerant ferromagnet was developed for small values of $q$ and $\omega$/$q$ and this theory tells that static susceptibility will follow CW law only for small values of $q$\cite{Moriya85}. It means that the value of $q$ is important in this theory. If we use primitive cell volume in place of Co atomic volume then the value of $q_B$ will be smaller and may fall in the region of $q$ for which CW law holds. It is necessary to do inelastic neutron scattering experiment in LaCoPO in order to evaluate the value of $\Gamma_0$. Using this value of $\Gamma_0$ and T$_0$ (obtained from present NMR results) it will be possible to determine $v_0$, which will verify whether Co atomic volume should be taken or the primitive cell volume in the expression for $q_B$.

\section{CONCLUSIONS}

The magnetic property of the itinerant ferromagnet LaCoPO was investigated in the oriented ($\overrightarrow{c}$$\parallel$$\overrightarrow{H_0}$) powder sample from $^{139}$La NMR Knight shift and the spin-lattice relaxation studies. The same measurements were also performed in random powder sample of LaFePO for comparison. The local magnetic field at the La nuclear site is isotropic and $T$ independent in LaFePO, while it is highly anisotropic and $T$ dependent in LaCoPO. The dominant contribution to this anisotropy arises from the electron nuclear hyperfine interaction, producing a negative spin polarization at the La electronic orbital, possibly through the RKKY interaction, while the dipolar contribution to the same is positive in sign and one order of  magnitude smaller. Moreover, the replacement of Fe by Co not only lowers the symmetry of the local magnetic field but also enhances the electric field gradient at the La nuclear site. This could arise due to some modification produced in the band structure, such that the contribution to the EFG from the non-$s$ electronic orbitals is enhanced along with a lowering in symmetry of the local magnetic field at the La nuclear site. Detailed band structure calculation showing the involvement of La electronic orbitals, is necessary to understand these findings. The $T$ dependence of $^{139}$La 1/$T_1T$ in LaCoPO shows the existence of non-negligible 3D spin fluctuations in both the paramagnetic and the ordered state over and above the dominant 2D spin fluctuations in the paramagnetic phase, observed from the previous $^{31}$P NMR relaxation data in the same oriented sample.\cite{Majumder09} The origin of this 3D spin fluctuations could be due to the existence of non-zero inter layer exchange interaction between the Co-P planes, which is  communicated through the La-O plane situated in between as shown in Fig.1. By applying the SCR theory of Moriya to the magnetization data of LaCoPO, we have also determined the different spin fluctuation parameters along with the same from La NMR relaxation data. The agreement was found to be quite satisfactory and suggests a more localized character of the 3d electrons in LaCoPO when compared with the same reported from the magnetization data in LaCoAsO. Finally, it is to be pointed out that the presence of $\sim$ 4\% La$_2$O$_3$ in the powder sample of LaCoPO should give rise to a signal at the reference position (42.37 MHz), which would be superimposed on the $^{139}$La central transition in LaCoPO at 300 K. However, below 300 K, presence of any signal from La$_2$O$_3$ (whose position is $T$ independent) should come out from the central line of LaCoPO. No such signal from La$_2$O$_3$ was detectable throughout the whole temperature range. Therefore, it would not affect the results.

\end{document}